\documentclass[pra,twocolumn,floatfix,nofootinbib]{revtex4-2}

\usepackage{braket}
\usepackage[utf8]{inputenc}
\usepackage[T1]{fontenc}
\usepackage{graphicx, placeins}
\usepackage{color}
\usepackage{comment}
\usepackage{amsmath}
\usepackage[unicode=true,pdfusetitle,bookmarks=true,bookmarksnumbered=true,bookmarksopen=false, breaklinks=false,pdfborder={0 0 0},pdfborderstyle={},backref=false,colorlinks=false, pdftitle={Degenerate Raman sideband cooling of $^{40}$K}]{hyperref}
\begin{document}

\title{Degenerate Raman sideband cooling of $^{40}$K atoms}

\author{Elad Zohar$^\dagger$, Yanay Florshaim$^\dagger$, Oded Zilberman, Amir Stern, and Yoav Sagi}

\email[Electronic address: ]{yoavsagi@technion.ac.il}

\affiliation{Physics Department and Solid State Institute, Technion --- Israel Institute of Technology, Haifa 32000, Israel}
\date{\today} 

\date{\today}
\begin{abstract}
We report on the implementation of degenerate Raman sideband cooling of $^{40}$K atoms. The scheme incorporates a 3D optical lattice, which confines the atoms and drives the Raman transitions. The optical cooling cycle is closed by two optical pumping beams. The wavelength of the laser beams forming the lattice is close to the D$_2$ atomic transition, while the optical pumping is operated near the D$_1$ transition. With this cooling method, we achieve temperature of $\sim$$1\mu$K of a cloud with $\sim$$10^7$ atoms. This corresponds to a phase space density of $\ge$$10^{-3}$. Moreover, the fermionic ensemble is spin polarized to conditions which are favorable for subsequent evaporative cooling. We study the dependence of the cooling scheme on several parameters, including the applied magnetic field, the detuning, duration, and intensity profile of the optical pumping beams. Adding this optical cooling stage to current Fermi gas experiments can improve the final conditions and increase the data rate.
\end{abstract}

\maketitle

\def\thefootnote{$\dagger$}\footnotetext{These authors contributed equally to this work}\def\thefootnote{\arabic{footnote}}

\section{Introduction}
Ultracold atomic gases are used in a wide range of applications, from studying many-body physics \cite{Bloch2008, Noh2016, Browaeys2020}, through quantum simulation and computation \cite{Ludlow2015, Endres2016, Cooper2018, Wang2020}, and all the way to precision metrology \cite{Cronin2009, Zhang2016, Wu2019}. Most such experiments require several cooling stages before the atoms can be used for the intended purpose. Laser cooling and evaporative cooling are the working horses of this preparation procedure. The former has the advantage of being efficient, namely, being relatively fast and incurring only minor loss of atoms. A natural limitation of laser cooling is the recoil energy an atom acquires following an emission of a single photon. To mitigate this effect, the cooling rate has to be reduced as the cooling process progresses. One way to achieve this is to design the process such that atoms will accumulate in a low energy final state which is almost decoupled from all the cooling lights. The buildup of population in this final dark state achieves another desirable goal -- spin polarization. The alternative route to preparing a sample in a well-defined internal state is by optical pumping, but it leads to heating due to spontaneous photon scattering. A cooling scheme that ends with all atoms being in a single dark state at a very low temperature solves this problem.

One of the most successful and widely used methods to approach and go below the recoil limit is to employ Raman transitions. Originally, the technique was developed in continuous space \cite{Kasevich1992}, with a series of velocity-selective Raman pulses. Later, it was adopted to a periodic potential, where it was called Raman sideband cooling (RSC) \cite{Hamann1998, Perrin1998}. In this scenario, the two Raman photons drive side-band transitions $n\rightarrow n-1$, where $n$ is the vibrational state in any given lattice site. Since the transition induced by a continuous Raman process is reversible, it cannot by itself lead to cooling. To break time-reversal symmetry, one can either make the Hamiltonian time-dependent (e.g., pulsing the Raman beams \cite{Kasevich1992}), or introduce a dissipative irreversible process, such as spontaneous emission. In the case of Raman cooling in an optical lattice, the latter is achieved by optical pumping. In the Lamb-Dicke regime, the spontaneous emission is not likely to change the vibrational state. Thus, the cooling cycle is closed with some of the vibrational energy removed.

A particularly clever implementation of Raman sideband cooling uses the lattice light itself to drive the Raman transitions \cite{Vuletic1998}. That means that the transition $n\rightarrow n-1$ needs to be degenerate, since the two Raman photons have the same frequency. Degenerate Raman sideband cooling (dRSC) was first realized with Cs atoms in a 1D geometry \cite{Vuletic1998}, and later extended to 3D \cite{PhysRevLett.84.439,Han2000}. It achieved atomic densities that approach the lattice site density and a temperature of $1.5$T$_\text{rec}$, where T$_\text{rec}=\hbar^2 k^2/m k_B$ is the recoil temperature. dRSC was successfully implemented with the bosonic isotopes $^{87}$Rb \cite{Hu2017, Urvoy2019}, $^{85}$Rb \cite{PhysRevA.97.023403}, and $^{39}$K \cite{Groebner2017}. To the best of our knowledge, dRSC was not implemented with fermionic atoms. Here, we report on a realization of dRSC with the fermionic isotope of potassium ($^{40}$K). Non-degenerate RSC was previously used to image single $^{40}$K atoms in an optical lattice \cite{Cheuk2015}. In our implementation, we start with a non-degenerate gas and  optimize the scheme to achieve the highest phase space density and prepare the gas for further cooling.

\begin{figure}
	\centering
	\includegraphics[scale=0.42]{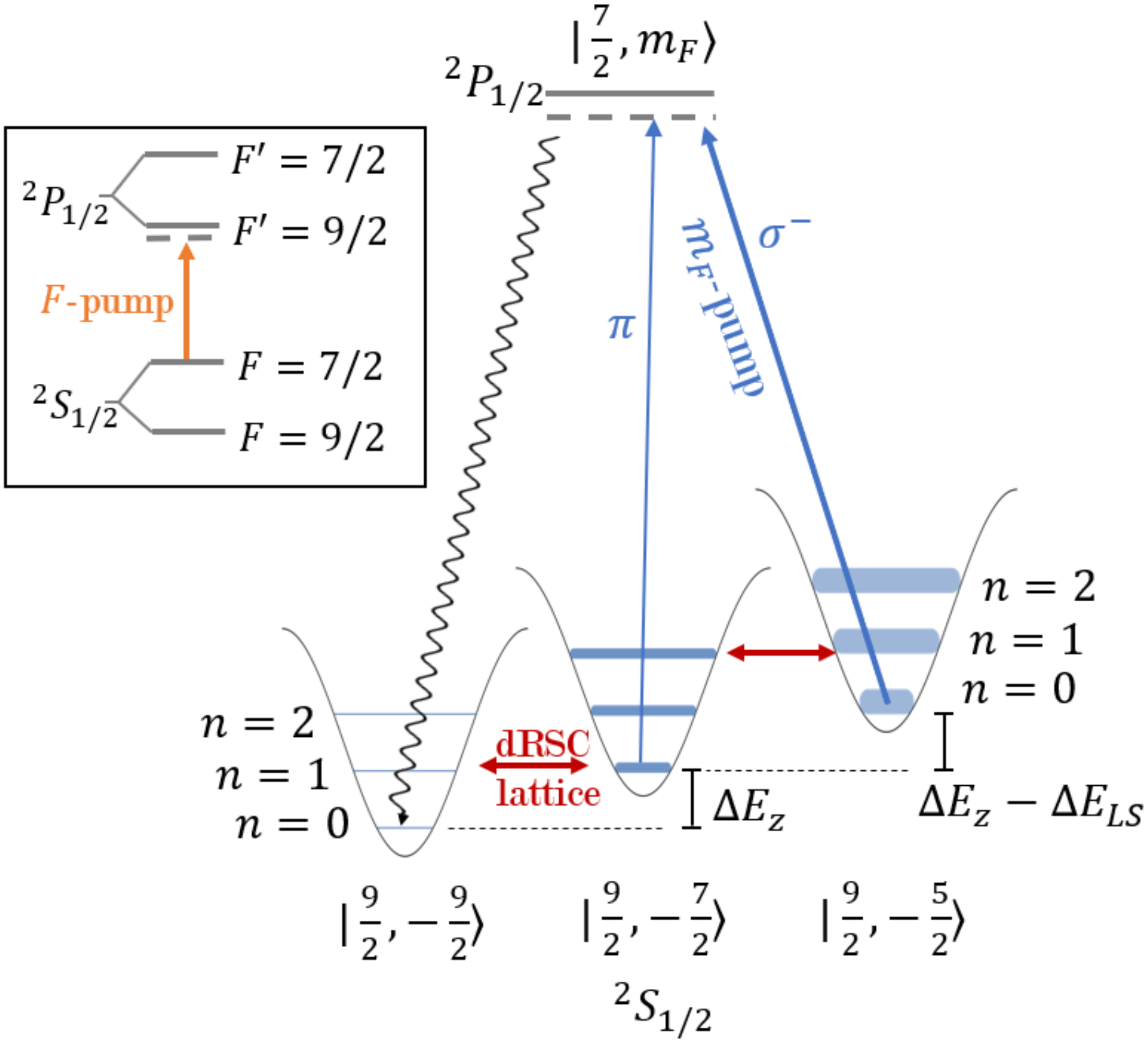}
	\caption{Degenerate Raman sideband cooling scheme with $^{40}$K atoms. The lattice provides the confining potential (here depicted for a single lattice site), as well as drives Raman transitions  (depicted by red horizontal arrows) between the different m$_F$ states of the $F=9/2$ ground state level with different vibrational principle numbers $n$. The energy of different m$_F$ states with the same $n$ is shifted due to the applied magnetic field. Necessary components in the cooling cycle are the m$_F$ and $F$ optical pumping beams. The former pumps atoms in the $F=9/2$ towards negative spin states, while the latter brings back into the cooling cycle atoms that decay to the $F=7/2$ level. The m$_F$-pump carries a strong $\sigma^-$ polarization component and a weak $\pi$ component. This leads to broadening of the $m_F\ge -5/2$ states (depicted by a larger width of the $n$ levels), as well as creates a differential light shift, $\Delta E_{\textrm{LS}}$, which, for red detuning shifts the m$_F=-5/2, -7/2$ to lower energies. The wavelengths of the Raman and optical pumping lasers are close to the D$_2$ and D$_1$ lines, respectively.}
	\label{fig:optical_transitions}
\end{figure}

Our complete cooling sequence begins with loading a 3D magneto-optical trap (MOT) with approximately $10^7$ atoms, followed by gray molasses cooling (GMC) \cite{Salomon2013, Fernandes2012} on the D$_1$ line. The GMC stage cools the atoms to $\sim$$5\mu$K, while leaving them occupying all spin states of the $F=9/2$ ground state manifold. Then, we load the cloud into the Raman optical lattice, and within a few milliseconds of dRSC we achieve a spin polarized sample with a temperature as low as $1\mu$K, which is around 1.2T$_\text{rec}$. The highest phase space density (PSD) we measure is around $10^{-3}$, two orders of magnitude higher than that of the cloud immediately after the GMC stage. The dRSC leaves over $80\%$ of the atoms in the $F=9/2, m_F=-9/2$ ground state, and the rest are at m$_F=-7/2$ (see Fig. \ref{fig:optical_transitions}). These conditions significantly improve the starting point and efficiency of subsequent forced evaporation, which we perform in a crossed far-off-resonance dipole trap. Since identical ultracold fermions cannot interact through symmetric s-wave scattering, the remaining $20\%$ of the atoms in the $m_F=-7/2$ state are essential for fast thermalization and efficient evaporation \cite{DeMarco1999, Geist2002}. Moreover, a mixture of these two spin states is a natural choice for quantum simulation experiments since they are stable against spin-exchange collisions and offer a convenient Feshbach resonance around $202.14$G \cite{Shkedrov2018}, which allows tuning the strength of their interaction \cite{Bohn2000}. It is also possible to change the spin composition by driving spin rotations with a radio-frequency radiation that matches the energy difference between the two spin states \cite{Shkedrov2022}.

\section{Experiment}

\begin{figure}
	\centering
	\includegraphics[scale=0.26]{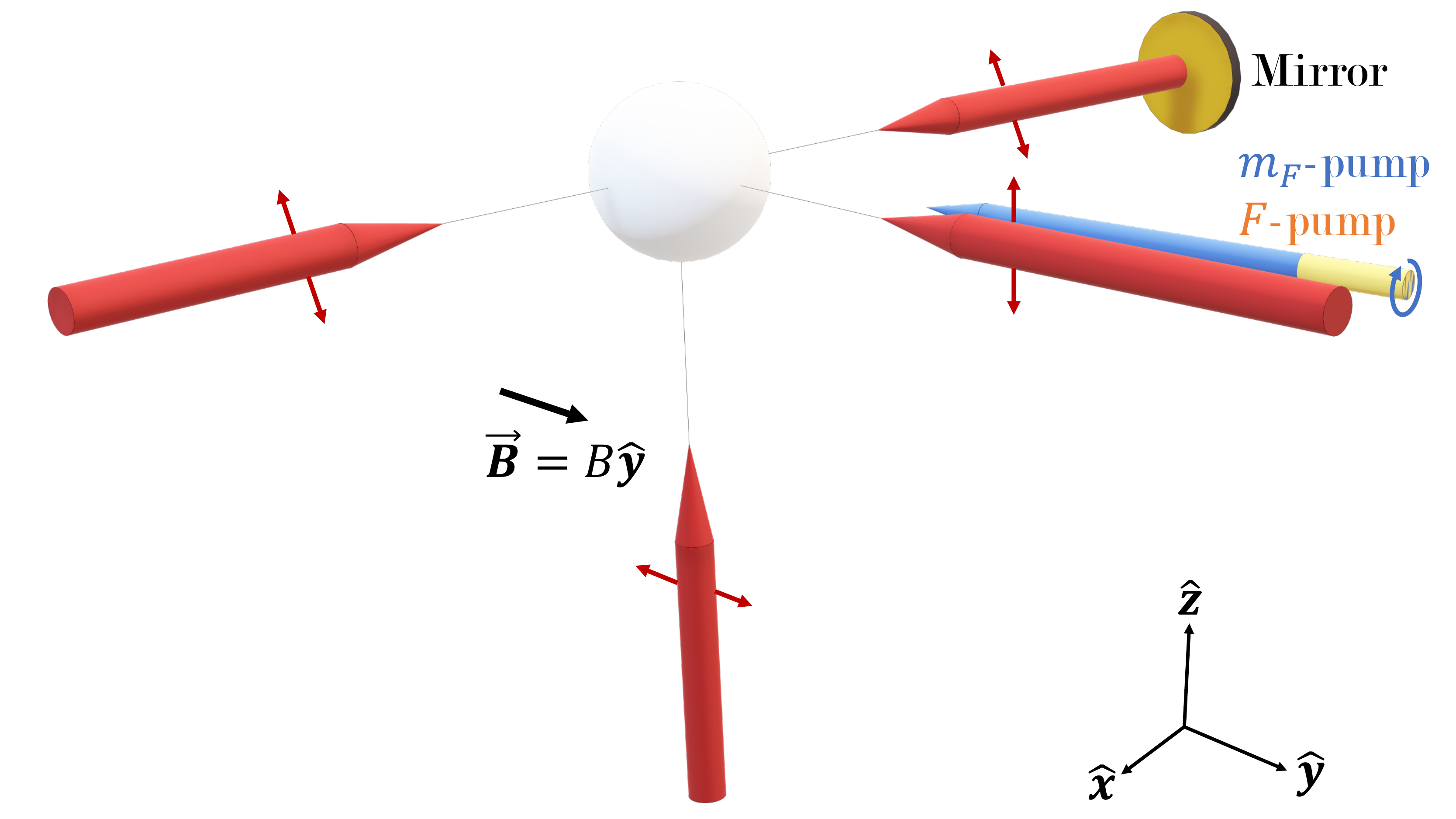}
	\caption{Schematic of the optical and magnetic setup of the dRSC scheme. The lattice is formed by four beams (thick red arrows). The two beams along the $\hat{x}$ direction are created by a single retro-reflected beam. The other two beams cross at an angle very close to $90^\circ$. The polarization of all lattice beams (depicted as thin arrows) lie in the $y$-$z$ plane. The magnetic field $\vec{B}$ is directed along the $\hat{y}$ direction. The circularly polarized m$_F$-pump and F-pump beams (blue and yellow arrow) propagate at an angle of approximately $5^\circ$ relative to the magnetic field, chosen to introduce a small $\pi$ polarization component required to complete the cooling cycle.}
	\label{fig:optical_setup}
\end{figure}

A sketch of the experimental setup is presented in Fig. \ref{fig:optical_setup}. The 3D lattice is created by four beams: one beam, propagating along the $x$-axis in the sketch, is retro-reflected, while the two others, propagating along the $y$ and $z$ axes, are free running. All beams are linearly polarized; The retro-reflected beam is polarized at an angle of $45^\circ$ relative to the $z$-axis, as depicted in Fig. \ref{fig:optical_setup}. This makes the dipole potential induced by the lattice isotropic in two out of three axes ($y$ and $z$). All polarizations of the lattice beams, as well as the magnetic field, lie in the same $y$-$z$ plane.

The source of the lattice beams is a distributed Bragg reflector (DBR) laser, whose frequency is red detuned $-11$ GHz with respect to the D$_2$ line. By deriving all lattice beams from the same source we ensure that relative phase shifts do not modify the lattice geometry, but only translate the entire lattice, which does not harm the cooling process \cite{Grynberg1993, Groebner2017}. The lattice laser is amplified by a home-built tapered amplifier (TA) and then passed through an acousto-optic modulator (AOM), for fast intensity control. It is also passed through a temperature stabilized solid etalon, to filter out the broadband amplified spontaneous emission (ASE). The etalon has a free spectral range (FSR) of $60$ GHz and a bandwidth of $1.6$ GHz.

After the etalon, we split the light once using a $\lambda/2$ plate and a polarizing beam splitter (PBS), and then a second time using a non-polarizing beam splitter (NPBS). The $\lambda/2$ plate allows us control over the ratio between the power of the retro-reflected beam ($\hat{x})$ and the $\hat{y}$ and $\hat{z}$ beams, which is typically set such that the former has twice the power of the latter. This ratio was optimized experimentally at a fixed lattice detuning. All beams are then sent to the experiment via three $5$ m single-mode polarization maintaining optical fibers. We sample and monitor the power of one of the beams with a photodiode at the fiber output, and then use a servo loop and the AOM before the fibers to stabilize it. The total power of all three beams is kept at $100$ mW with a relative stability of $\sim 10$ ppm, which yields a peak intensity of $1590\,\text{mW}/\text{cm}^2$ ($795\,\text{mW}/\text{cm}^2$) and thus a calculated depth of 586$\mu $K (238$\mu $K) in the $\hat{x}$ ($\hat{y}$ and $\hat{z}$) direction. With these parameters, the calculated harmonic trapping frequencies in each of the lattice sites are $344$ kHz in the $\hat{x}$ direction and $138$ kHz in the $\hat{y},\;\hat{z}$ directions.
The magnetic field is applied using two pairs of external Helmholtz coils, such that the total field points in the desired direction.

There are two optical pumping beams in our setup: m$_F$-pump, and F-pump. The m$_F$-pump is set as close as possible to perfect circular polarization. Due to a small angle between it and the magnetic field, it carries a strong (weak) $\sigma^-$ ($\pi$) component. The strong $\sigma^-$ component pumps the atoms to the $|F=\frac{9}{2}, m_F=-\frac{7}{2}\rangle$ state. Because the entire process is performed in the Lamb-Dicke regime, that is, the Lamb-Dicke parameter $\eta=\sqrt{\frac{E_R}{\hbar\omega_{\textrm{lattice}}}}=0.31\; (0.49)$ for the $\hat{x}$ ($\hat{y}$ and $\hat{z}$) direction ($E_R$ is the photon recoil energy), this process is unlikely to change the vibrational state of the atom. This means that atoms undergo several cycles which consist of a Raman transition that lowers the vibrational state and an m$_F$-pump which resets the atoms to the m$_F=-7/2$, until they reach the ground state. The weak $\pi$ component is used to pump the atoms to m$_F=-9/2$, at which point they are decoupled from both the Raman and optical pumping processes. The second optical pumping beam (F-pump) is used to pump atoms which decay from the m$_F$-pumping process to the $F=7/2$ ground state manifold, which is not part of the cooling process. The F-pumping is performed through the $F=9/2$ state of the D$_1$ level. We send both optical pumping lights through the same optical fiber. The m$_F$-pump is locked to the D$_1$ line using a saturated absorption spectroscopy (SAS) setup \cite{PhysRevLett.23.631}, and the F-pump frequency is stabilized relative to the m$_F$-pump by monitoring their beat signal \cite{Schuenemann1999}. The required frequency shifts which bring each of the optical pumping lights to the desired detunings are achieved using several AOMs.

The full loading sequence is thus as follows. $^{40}$K atoms are dispensed from home-built dispensers, then trapped and cooled in a 2D MOT \cite{Bhushan:17, NELLESSEN1990300} and pushed into the main vacuum chamber. There, a 3D MOT captures and cools them further during $6$ seconds of loading time. Following the 3D MOT, the atoms are cooled for $8$ ms with a gray molasses D$_1$ cooling scheme \cite{Salomon2013, Fernandes2012}, during which the magnetic field is set to zero using three pairs of Helmholtz coils. We then turn off the D$_1$ cooling light and simultaneously ramp up in $0.8$ ms the Raman lattice power and the magnetic field. The magnetic field is ramped to $500$ mG using two of the three Helmholtz coils, such that it points in the desired direction, as depicted in Fig. \ref{fig:optical_setup}. With the lattice and magnetic fields ready, we turn on the m$_F$-pump and F-pump beams. During optimization, we found that optimal cooling is achieved when the m$_F$-pump is linearly ramped up from a low initial intensity of $0.1$ mW/cm$^2$ to $2.1$ mW/cm$^2$ during the $10$ ms of dRSC. We discuss this observation in the results section. 

After $10$ ms of cooling, we turn off the optical pumping lights and ramp down the lattice for another $0.8$ ms. The next step in our experiment is loading the cloud to a crossed dipole trap. We find no significant difference in cooling efficiency if the crossed trap is kept on during the GMC and dRSC processes, and this is indeed what we do. Once the lattice is ramped down, we capture more than $10^6$ atoms (12\% of the atoms in the dRSC) in the crossed dipole trap. Since this process compresses the atoms in real space, and the phase space density is preserved, the temperature immediately after their loading is $7\mu $K. The results presented below, however, were taken without the crossed dipole trap, therefore they give a direct characterization of only the dRSC process.

\section{Results}

\begin{figure}
	\centering
	\includegraphics[scale=0.52]{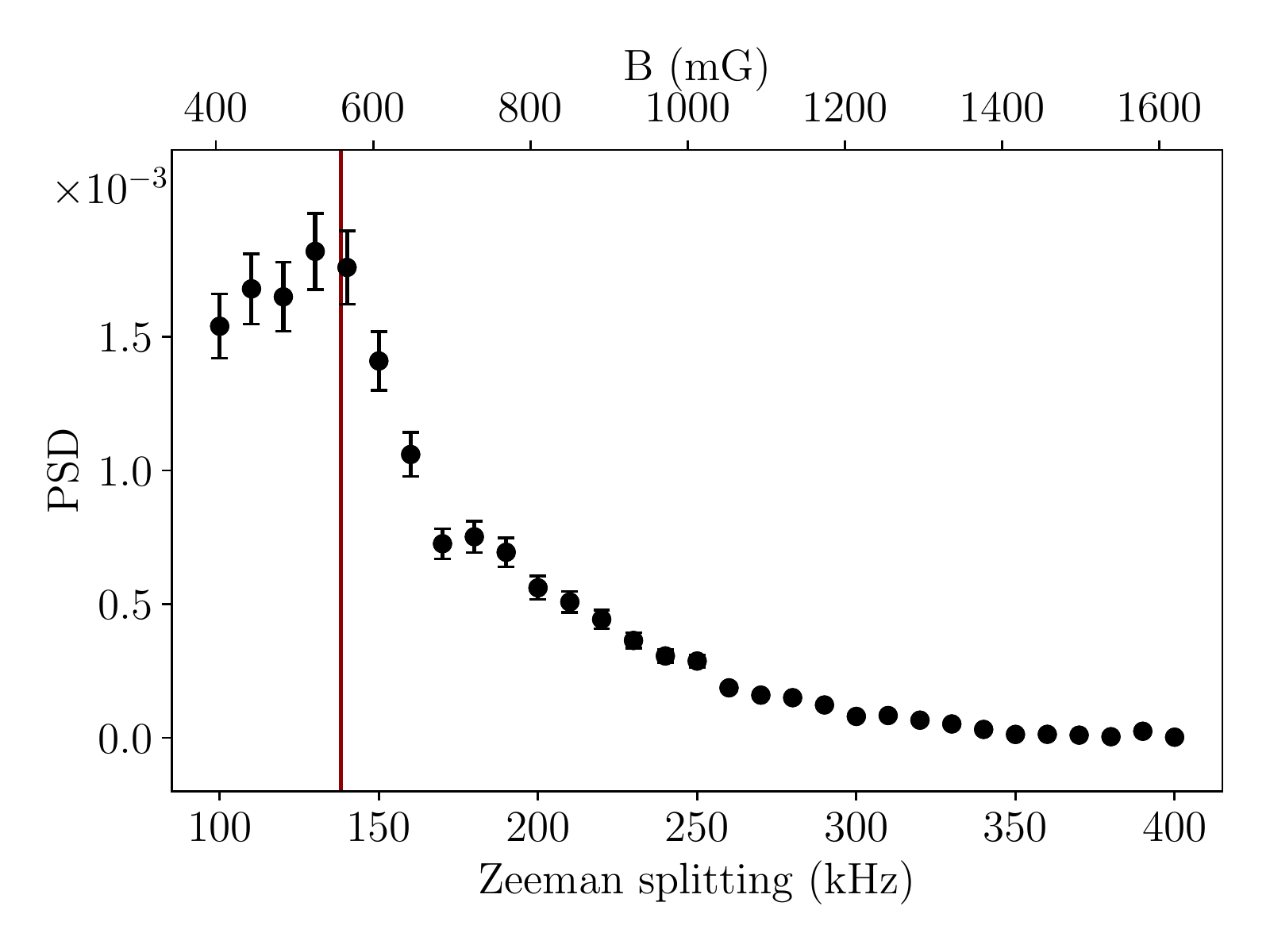}
	\caption{The final phase space density as a function of the applied magnetic field. The upper x-axis provides the actual field magnitude, while the bottom x-axis gives the Zeeman splitting between the $|9/2,-9/2\rangle$ and $|9/2,-7/2\rangle$ states associated with this field. Since we are working in the linear regime of the Zeeman splitting, other m$_F$ levels would be shifted approximately (up to a few Hz) by the same amount. The vertical red line marks the harmonic trapping frequency in the $\hat{y},\;\hat{z}$ directions. Error bars in this paper represent one standard deviation.}
	\label{fig:magnetic_field}
\end{figure}

To characterize the gas conditions, we measure the integrated density distribution of the cloud using absorption imaging, both \emph{in situ} and after a time-of-flight of typically $30$ ms. We fit the distributions measured at the two times with a Gaussian, and from the change in their widths we extract the cloud temperature. In addition, integration over the density gives us the total number of atoms. We can then calculate the PSD given by $n\lambda_\text{db}^3$, where $n$ is the density and $\lambda_\text{db}$ is the thermal de-Broglie wavelength. We use the number of atoms, and the volume as approximated by a sphere of a radius $\frac{\sigma_x+\sigma_y}{2}$, where $\sigma_i$ is the \emph{in situ} Gaussian width in direction $i$, to calculate $n$.

We now turn to characterize the dependence of the cooling sequence on various parameters. First, we calibrate the optimal magnetic field magnitude required for cooling. To this end, we fix all other parameters, scan the magnetic field, and extract the PSD. As can be seen in Fig. \ref{fig:magnetic_field}, The optimum fits the theoretical lattice frequency in the two isotropic directions. There is no distinct peak corresponding to the trapping frequency in the $\hat{x}$ direction because the Raman coupling in that direction is vanishingly small due to the standing wave \cite{Vuletic1998}.

\begin{figure}
\centering
\includegraphics[scale=0.52]{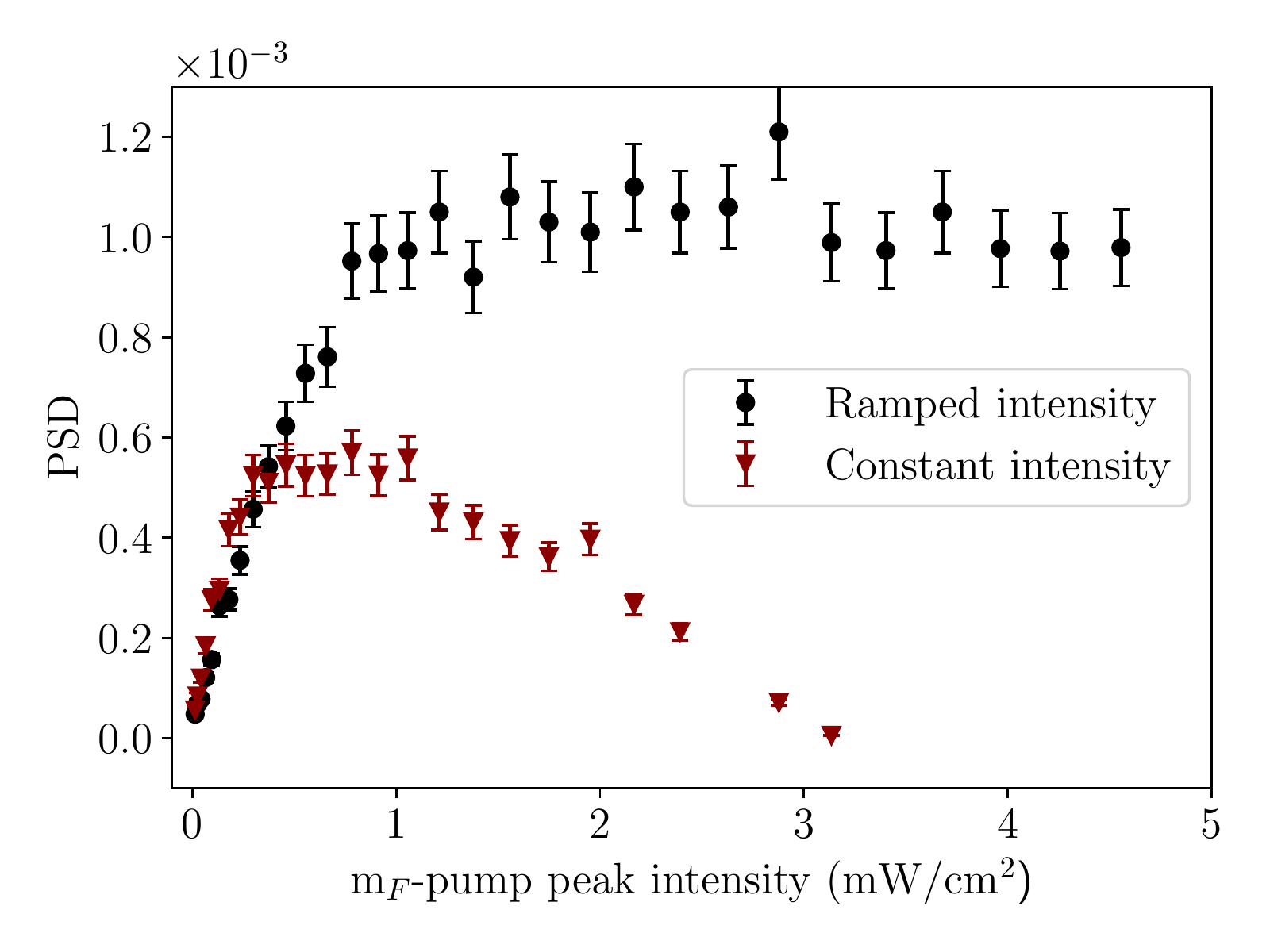}
\caption{Phase space density for different m$_F$-pump intensities. 
Runs with constant (ramped) pump intensity are marked by triangles (circles). In the ramped intensity experiments, the x-axis gives the final m$_F$-pump intensity after a linear ramp from an initial intensity of 0.1 mW/cm$^2$. We attribute the improvement when ramping the intensity to a smooth change of the light shift induced by the m$_F$-pump, which compensates for the inhomogeneous anharmonic broadening of the transition frequencies in the lattice.}
\label{fig:mf_power}
\end{figure}

The second parameter we scan to optimize the dRSC is the m$_F$-pump intensity, with the detuning fixed at $-8$ MHz. The result is presented in Fig. \ref{fig:mf_power}. Working with constant intensity, we observe an initial improvement of the PSD with increasing optical pumping intensity. The PSD saturates around $0.5\times 10^{-3}$ for intensity in the range of $0.5-1$ mW/cm$^2$. At even higher intensity, the PSD decreases. We attribute this behavior to the interplay between the Raman and optical pumping transition rates. When the latter is much higher than the former, transitions that increase the vibrational quantum number become too frequent, and atoms eventually become untrapped and leave the lattice. This is particularly detrimental when there is a substantial occupation of atoms at $m_\text{F}>-5/2$ states, as is the case immediately at the beginning of the dRSC process, following the GMC stage.

Instead, we find that it is beneficial to initiate the optical pumping at low intensity ($0.1$ mW/cm$^2$) and then ramp it linearly to a higher value. Results following this procedure are shown in Fig. \ref{fig:mf_power} versus the final intensity. With ramping we obtain a two-fold improvement in the final PSD. There are two plausible explanations for this effect. First, a higher efficiency of the optical pumping with the $\pi$ component at the final stages of the cooling process. Second, the linear ramp induces a smooth change of the light shift ($\Delta E_\text{LS}$ in Fig. \ref{fig:optical_transitions}), which is better suited to the non-uniform and anharmonic lattice potential. In what follows, we fix the intensity profile of the m$_F$-pump to linear ramping with a final intensity of $2.1$ mW/cm$^2$.

\begin{figure}
\centering
\includegraphics[scale=0.52]{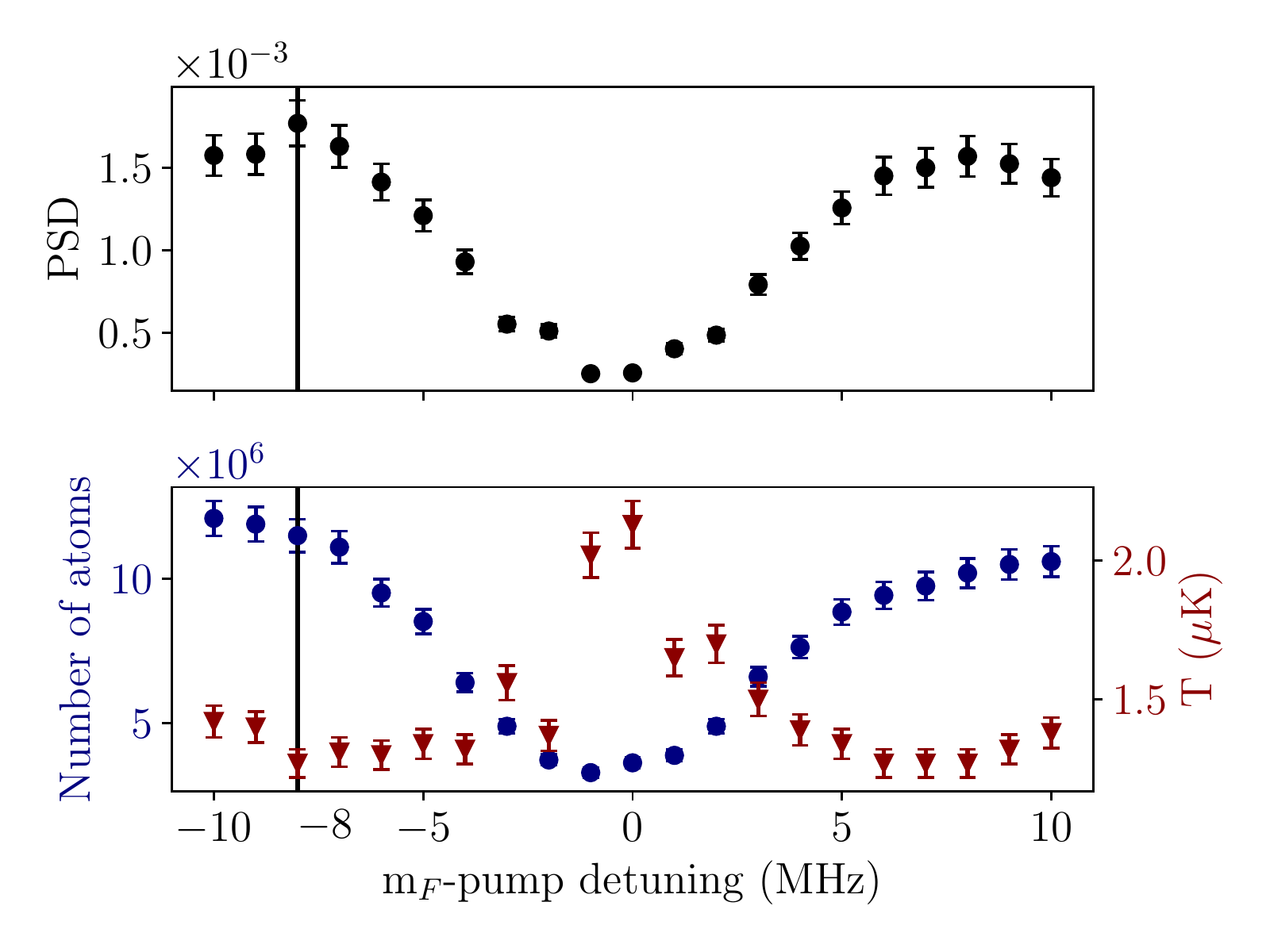}
\caption{Upper panel: the phase space density after applying dRSC with different m$_F$-pump detuning. It exhibits a slightly better performance on the red side. Lower panel: number of atoms (blue circles, left axis) and temperature (red triangles, right axis) from which the PSD was calculated. The chosen working condition is marked by a vertical black line.}
\label{fig:mf_det}
\end{figure}

Next, we study the dependence of the cooling process on the m$_F$-pump detuning (Fig. \ref{fig:mf_det}). Previous implementations of dRSC with other atomic species found that it is advantageous to work with a blue detuned m$_F$-pump \cite{PhysRevLett.84.439, Groebner2017}. In contrast, we observe that the dependence is almost symmetric with respect to the sign of the detuning, with slightly better results obtained when working with a red detuned light. As for the F-pump, we find that the detuning and power do not affect the final temperature, or increase the loss, as long as the intensity is above $4$ mW/cm$^2$ and the detuning is at least $-4$ MHz. This behavior is inline with previous results obtained with $^{39}$K  \cite{Groebner2017}.

\begin{figure}
\centering
\includegraphics[scale=0.52]{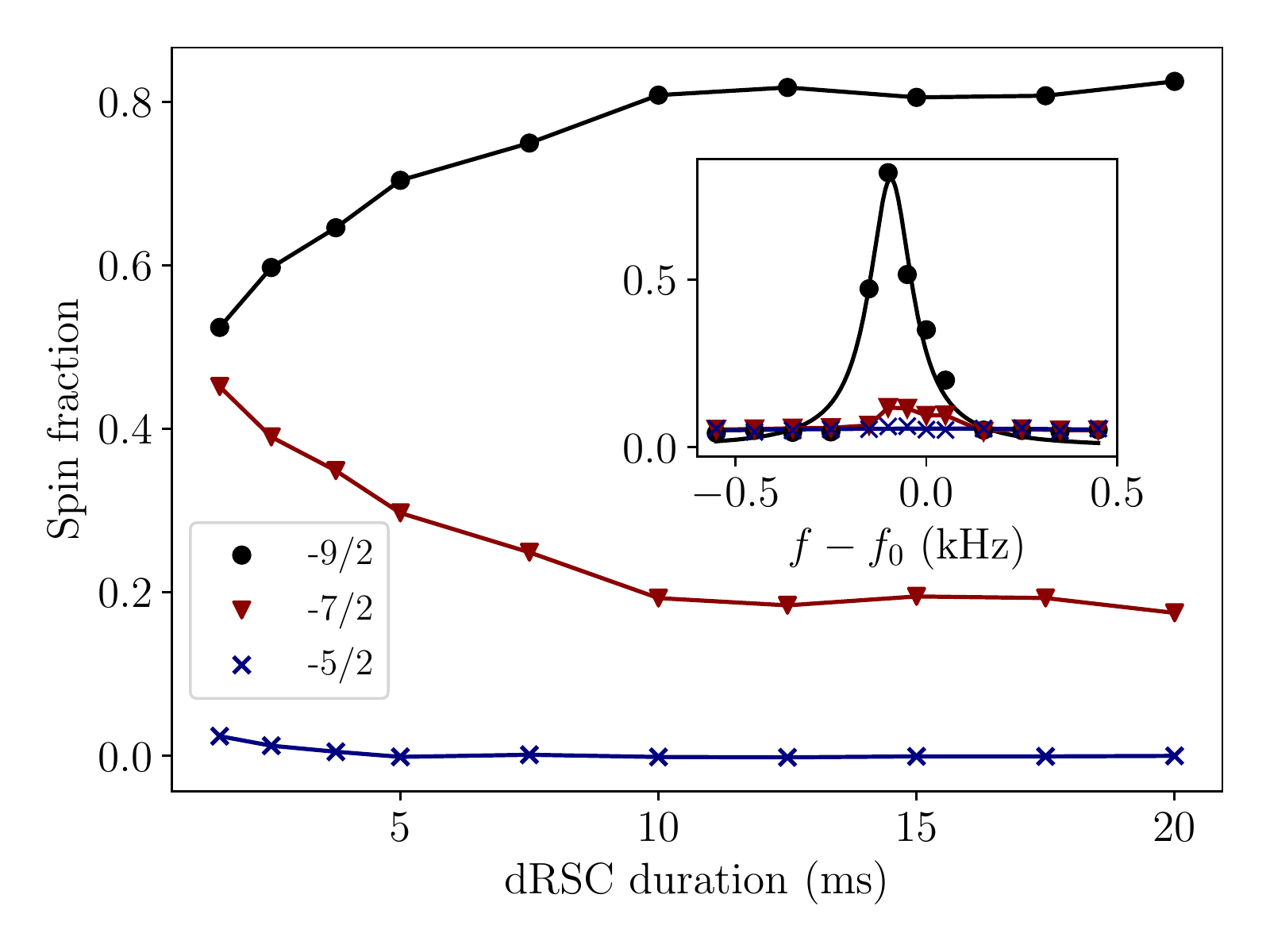}
\caption{Fraction of different m$_F$ states versus the duration of the dRSC sequence. Atoms are quickly pumped to occupy the two lowest m$_F$ states, after which the small $\pi$-component of the m$_F$-pump continues to reduce the population in the m$_F=-7/2$ state. The inset shows an example of the spectrum from which the fractions are extracted, where all three resonance frequencies of the different m$_F$ states are centered around the bare transition resonance, $f_0$, which depends on m$_F$ and the magnetic field. The spectrum is fitted with a Lorentzian to find the spectral weight.}
\label{fig:spin_comp}
\end{figure}

One of the advantages of dRSC is its ability to spin polarize the ensemble. To determine the final spin composition, we perform microwave (MW) spectroscopy after the cooling process is done. First, we load the atoms from the lattice to a crossed optical dipole trap. Then, we apply a MW field whose frequency we ramp over $50$kHz to compensate for magnetic field fluctuations. The MW field drives a ${|F=\frac{9}{2}, m_F\rangle \rightarrow |F=\frac{7}{2}, m_F'\rangle}$ transition in the ground state manifold. The MW central frequency determines which $m_F, m_F'$ levels we address, based on the Zeeman splitting. Then, we turn on a quadruple magnetic field which traps only the atoms transferred by the MW pulse while the other atoms are ejected out \cite{Shkedrov2018}. Finally, we detect the atoms captured in the quadruple trap using MOT fluorescence. The inset of Fig. \ref{fig:spin_comp} shows an example of this spectroscopy, from which we extract the relative population in each state. In the main graph of Fig. \ref{fig:spin_comp} we present the spin composition for different durations of the dRSC. One can see that after a few milliseconds, over 95\% of the atoms occupy the two m$_F=-9/2,-7/2$ states. In about $10$ms, their fraction saturate towards $80\%$ and $20\%$, respectively. In an ideal dRSC, the m$_F=-9/2$ population should ultimately approach unity. In practice, off-resonant excitations to the $F'=9/2$ state, imperfect polarization of the $m_F$-pump, and single photon scattering from the Raman lattice beams contribute to the observed asymptotic spin composition. 

\begin{figure}
\centering
\includegraphics[scale=0.5]{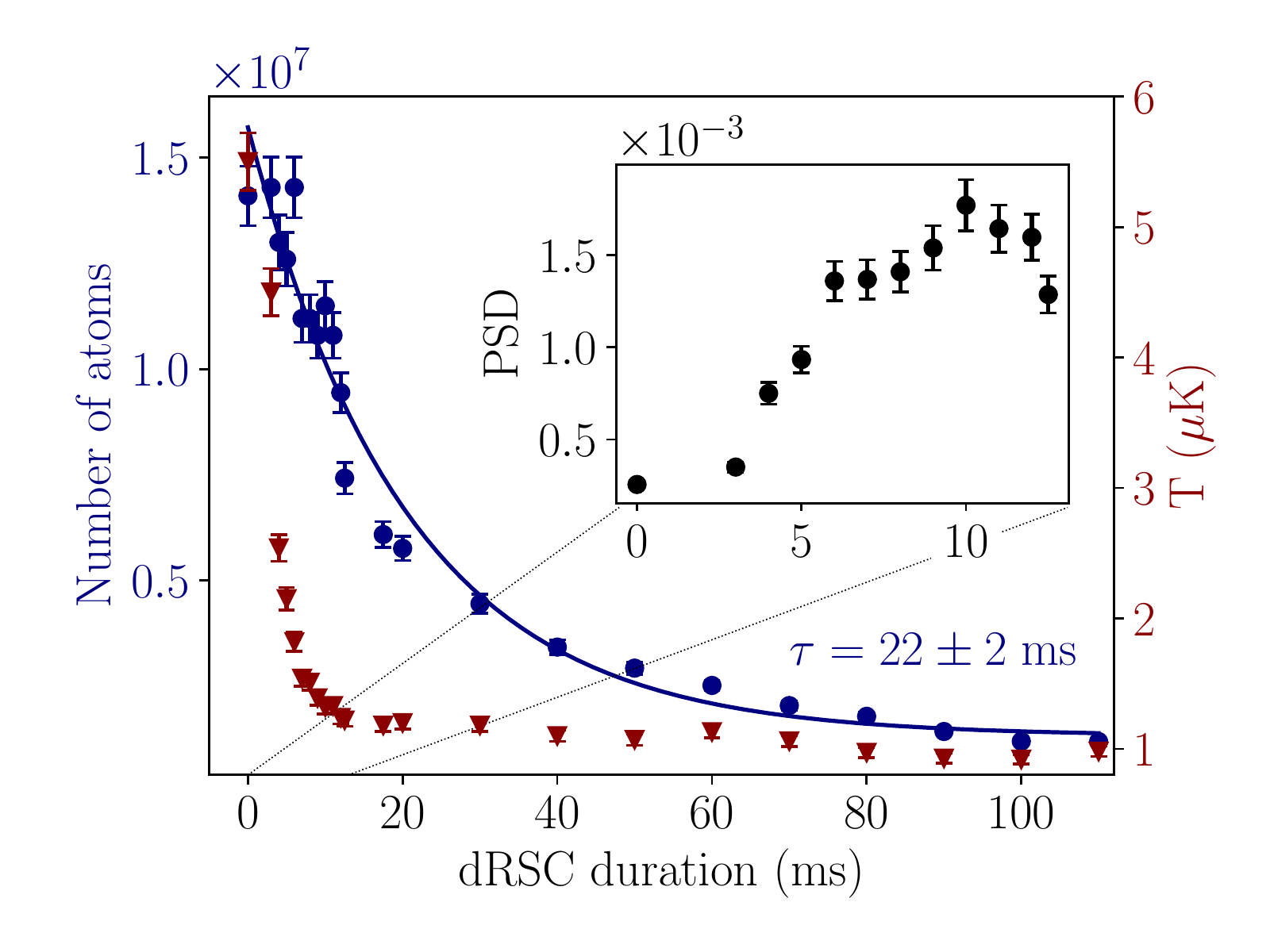}
\caption{
Number of atoms (circles, left axis) and temperature (triangles, right axis) as a function of the duration of the dRSC. For measurements shorter than or equal to $10$ ms, the intensity of the m$_F$-pump is linearly ramped between the same initial and final values during the dRSC duration. For longer times we keep all lights at their final intensity from 10 ms and on. The number of atoms is fitted with an exponential decay, yielding $\tau=22\pm2$ ms. The inset shows a close up of the phase space density in the first 12 ms. The point at zero time represents the conditions after the GMC. The asymptotic temperature we observe after 10 ms of cooling is $\sim$$1 \mu$K, and at this point the PSD peaks at $\sim$$1.5\times 10^{-3}.$}
\label{fig:lifetime}
\end{figure}

Finally, we measure the lifetime of the atoms in the lattice. This measurement is done in the following way: the first $10$ ms are performed in the same manner as described before, including ramping the m$_F$-pump. For longer durations, we keep all lights and the magnetic field at a fixed value after the initial $10$ ms. The resulting number of atoms and temperature versus the total duration are shown in Fig. \ref{fig:lifetime}. The inset shows the PSD in the first $12$ ms. Based on this measurement, we choose the cooling time to be $10$ ms. We fit the number of atoms data with an exponentially decaying function and find a timescale of $\tau=22\pm2$ ms, much shorter than the vacuum limited lifetime of $6.5\pm0.5$ sec. Since the cooling achieves optimal conditions in $10$ ms, this lifetime does not harm the usefulness of the dRSC technique. We attribute the loss to the spatial Gaussian profile of the lattice beams, which gives rise to a spatially varying lattice depth and harmonic trapping frequency. Specifically, the lattice beams' waist radius is around $1$ mm while the atomic cloud radius is around $0.5$ mm. This means that in the outer regions of the lattice the Raman degeneracy condition is not met and atoms are heated and lost. We calculate the heating rate due to single photon scattering by the lattice beams to be $\sim$$2 \mu$K/ms, which is of the same order as the cooling rate. The loss rate could be reduced by increasing the lattice detuning and working with larger lattice beams. Both of these solutions, however, require considerably more laser power. 

\section{Summary and Outlook}
We have presented a first experimental implementation of dRSC with the fermionic isotope of potassium $^{40}$K. The cooling scheme achieves a six fold reduction in temperature, compared to gray molasses cooling, and concurrently spin polarizes the sample. By ramping the m$_F$-pump light intensity up, we achieve a two-fold improvement in phase space densities achieved by this scheme when compared to using a constant intensity. The dRSC scheme greatly improves the loading conditions to a crossed dipole trap for subsequent evaporative cooling. Our implementation employs an optical lattice close to the D$_2$ line and optical pumping beams red detuned relative to the D$_1$ line. In the future, this separation of wavelengths can allow sensitive detection of the atoms position using the dRSC scheme and filtering out efficiently unwanted D$_2$ light scattering.

\begin{acknowledgments}
This research was supported by the Israel Science Foundation (ISF), grants No. 3491/21 and 1779/19, and by the Pazy Research Foundation.
\end{acknowledgments}
%


\end{document}